# Simulations with Complex Measures

T D Kieu and C J Griffin[a]

[a]School of Physics, University of Melbourne, Parkville Vic 3052, Australia

Towards a solution to the sign problem in the simulations of systems having indefinite or complex-valued measures, we propose a new approach which yields statistical errors smaller than the crude Monte Carlo using absolute values of the original measures. The 1D complex-coupling Ising model is employed as an illustration.

## 1. The sign problem

The simulation of many interesting statistical and quantum systems suffers from the sign problem when the generating function measures are indefinite or are complex [1]. Some important examples are Lattice QCD with finite temperature and density and chiral gauge theory.

If the measure $\rho(x)$ of a generating function suffers from the sign fluctuation then another positive definite function $\tilde{\rho}(x)$ must be chosen for the Monte Carlo (MC) evaluation of the expectation value of an obsevable $\Theta$,

$$\langle \Theta \rangle = \int_x (\Theta(x)\rho(x)/\tilde{\rho}(x))\tilde{\rho}(x) \Big/ \langle\langle 1 \rangle\rangle, \quad (1)$$

$$\langle\langle 1 \rangle\rangle \equiv \int_x (\rho(x)/\tilde{\rho}(x))\tilde{\rho}(x), \quad (2)$$

In general it is desirable to choose $\tilde{\rho}$ independently of $\Theta$, and we will concentrate on the estimate of the denominator $\langle\langle 1 \rangle\rangle$ because of its appearance in all the measurements. It is a simple variational problem to show that the MC probability density, properly normalised, which minimises the variance of $\langle\langle 1 \rangle\rangle$ which is $\sigma/\sqrt{\#\text{independent configurations}}$ and where

$$\sigma^2 = \int_x |\rho(x)/\tilde{\rho}(x) - \langle\langle 1 \rangle\rangle|^2 \, \tilde{\rho}(x), \quad (3)$$

must be

$$\tilde{\rho}(x) = |\rho(x)| \Big/ \int_x |\rho(x)| \, . \quad (4)$$

Thus the sign of $\rho(x)$ is now treated as part of the quantity whose expectation is to be measured; hence the name average sign approach [1].

However, when the denominator $\langle\langle 1 \rangle\rangle$ is vanishingly small, $\sigma^2$, though minimised, is $\approx 1 \gg \langle\langle 1 \rangle\rangle$ since $\rho/|\rho| = \pm 1$. Then the evaluation of $\langle \Theta \rangle$ becomes unreliable unless the number of independent configurations is many orders of magnitude greater than the large number $1/\langle\langle 1 \rangle\rangle^2$. The fluctuation of sign of the measure over configuration space thus renders ineffective the sampling guided by this crude MC method (4). This is the content of the sign problem.

Many approaches have been proposed to tackle the sign problem but none is satisfactory [2]. In the next section we will improve on the one presented above.

## 2. A new approach

We can write the generating function as integrals over two configurational subspaces, $\int_x \rho(x) = \int_X \int_Y \rho(X,Y)$, in such a way that the multi-dimensional integral over $Y$ can be evaluated analytically or well approximated:

$$\varrho(X) = \int_Y \rho(X,Y). \quad (5)$$

As in the last section, one can easily prove that the MC weight $\tilde{\varrho}$ that minimises

$$\sigma'^2 = \int_X |\varrho(X)/\tilde{\varrho}(X) - \langle\langle 1 \rangle\rangle|^2 \, \tilde{\varrho}(X), \quad (6)$$

is

$$\tilde{\varrho}(X) = |\varrho(X)| \Big/ \int_X |\varrho(X)| \, . \quad (7)$$

It then follows that the variance for this new weight is not bigger than that for $\rho(X,Y)$.

$$\sigma'^2 - \sigma^2 =$$

$$\int_X |\varrho(X)|^2/\tilde{\varrho}(X) - \int_X \int_Y |\rho(X,Y)|^2/\tilde{\rho}(X,Y),$$
$$= \left(\int_X |\varrho(X)|\right)^2 - \left(\int_X \int_Y |\rho(X,Y)|\right)^2, \quad (8)$$
$$= \left(\int_X \left|\int_Y \rho(X,Y)\right|\right)^2 - \left(\int_X \int_Y |\rho(X,Y)|\right)^2,$$

the second line follows from (4) and (7); the last line is from (5) and always less than or equal to zero because of the triangle inequality. The equality occurs if and only if $\rho(X,Y)$ is semi-definite (either positive or negative) for each $X$-configuration. In particular, when there is no sign problem in the first place, expression (6) yields the same statistical deviation as the crude one.

To deal with complex integrands, of which indefinite measures are special cases, we adopt the definition (3) of the variance extended to the absolute values of complex numbers. Statistical analysis from this definition is the same as the standard analysis; except that the range of uncertainty should now be depicted as the radius of an 'uncertainty circle' centred on some central value in the complex plane. All the manipulations above remain valid.

Our approach is now clear. The measures are first summed over a certain subspace, the integration (5) above, to facilitate some partial phase cancellation. Absolute values of these sums are then employed as the MC sampling weights (7). The numerator in (1) can be obtained from an appropriate derivative of the generating function after the partial summation.

## 3. Complex coupling Ising spins

Owing to the nearest-neighbour interactions, the lattice can be partitioned into odd and even sublattices, of which the Ising spins $s_i (= \pm 1)$ on site $i (\in$ the sublattice) do not interact with each other. Absolute values of sums of the complex-valued weights over the even sublattice, say, are the new Monte Carlo weights.

In our simulations, periodic boundary condition is imposed on the 1D chains of 128 Ising spins which become 64 spins after the partial summation. Ensemble averages are taken over 1000 configurations, out of $2^{128}$ possible configurations.

They are separated by 30 heat bath sweeps which is sufficient for thermalisation and decorrelation in all cases except perhaps one, as will be demonstrated shortly. A heat bath sweep is defined to be one run over the chain, covering each spin in turn. The numbers of trials per sweep are different before and after the partial summation because the numbers of spins to be updated are not the same. Both hot and cold initial configurations are used and will be indicated when needed. The crude MC weights of the average sign method are

$$\tilde{\rho}(\{s\}) \sim \prod_{\text{all sites}} \exp(J_{\text{re}} s_i s_{i+1}) \quad (9)$$

and the improved MC weights are

$$\tilde{\varrho}(\{s\}) \sim \prod_{\text{odd sites}} |\cosh(J(s_i + s_{i+1}))|, \quad (10)$$

with coupling $J = (J_{\text{re}}, J_{\text{im}})$ and no external field.

We have presented some measurements in the table by exact, improved MC and crude MC methods respectively. Expressions for the magnetisation and susceptibility are obtained from appropriate derivatives of corresponding partition functions. The complex correlation length $\xi$ is calculated from the transfer matrix eigenvalues.

The autocorrelation of the unit operator in Fig. 1 show that the noise is too overwhelming in the crude simulation to tell whether 30 sweeps are sufficient for thermalisation or not. In contrast, the improved simulation is very well-behaved.

## 4. Conclusions

Owing to the sign cancellation in the partial sums, our approach can offer substantial improvements over the crude average sign method and may work even when the later fails, in the region of long correlation length and vanishing partition function.

The choice for splitting of the integration domain is arbitrary and its effectiveness depends on the physics of the problem. If the interactions are short-range (not necessary nearest-neighbour), maximal, non-interacting sublattices can always be chosen to provide a natural splitting, which is the particular splitting for our illustrative example. The approach of Mak in [2] could be considered as yet another particular and non-trivial splitting where the inner integration was approximated in a certain manner.

| Coupling | | Magnetisation per spin | Susceptibility per spin | $\xi^{-1}$ |
|---|---|---|---|---|
| (0.1,0.1) | exact | (0,0) | (1.1971,0.2427) | (1.96,0.78) |
| | improved | (0.0007,0.0000) [0.0024] | (1.1921,0.2459) [0.0356] | |
| | crude | (-0.0032,0.0016) [0.0062] | (1.1651,0.2883) [0.1460] | |
| (0.01,0.1) | exact | (0,0) | (0.9999,0.2027) | (2.29,1.47) |
| | improved | (-0.0006,-0.0001) [0.0021] | (1.0481,0.2231) [0.0268] | |
| | crude | (0.0023,-0.0080) [0.0055] | (1.0868,0.1374) [0.1198] | |
| (0.01,0.5) | exact | (0,0) | (0.5512,0.8585) | (0.60,1.55) |
| | improved | (-0.0031,-0.0047) [0.0027] | (0.5604,0.8163) [0.0428] | |
| | crude | (-0.0522,0.0820) [0.2074] | (0.6808,5.4507) [9.2788] | |
| | | (0.0123,-0.0723) [0.0830] | (1.5323,1.3885) [2.3122] | |

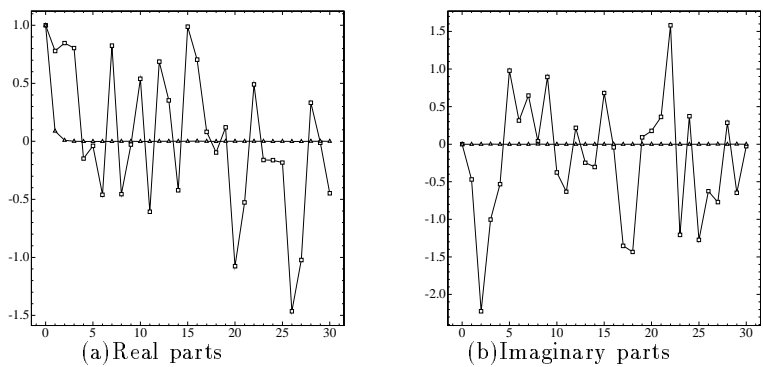

(a)Real parts      (b)Imaginary parts

Figure 1. The autocorrelation versus the number of sweeps for $J = (0.01, 0.5)$ from a total of $2 \times 10^6$ configurations. The boxes are from crude weights; triangles, improved weights.

**Acknowledgements**

We wish to thank Chris Hamer, Bruce McKellar, Mark Novotny and Brian Pendleton for discussions and support. TDK acknowledges the support of a Research Fellowship from the Australian Research Council.